\documentclass[pdftex, pra, twocolumn, showpacs, floatfix, superscriptaddress]{revtex4}

\usepackage{times}
\usepackage{amsfonts}
\usepackage{amssymb}
\usepackage{amsmath}
\usepackage{graphicx}
\usepackage{color}
\definecolor{darkblue}{rgb}{0, 0, 0.7}
\usepackage[colorlinks=true, breaklinks=true, linkcolor=darkblue, citecolor=darkblue, urlcolor=darkblue]{hyperref}

\newcommand{\doilink}[2]{\href{http://dx.doi.org/#1}{#2}}

\begin{document}

\title{Ground-state properties of few dipolar bosons in a quasi-one-dimensional harmonic trap}

\author{F. Deuretzbacher}
\email{fdeuretz@itp.uni-hannover.de}
\affiliation{Institut f\"ur Theoretische Physik, Leibniz Universit\"at Hannover, Appelstrasse 2, D-30167 Hannover, Germany}

\author{J. C. Cremon}
\affiliation{Mathematical Physics, LTH, Lund University, Post Office Box 118, S-22100 Lund, Sweden}

\author{S. M. Reimann}
\affiliation{Mathematical Physics, LTH, Lund University, Post Office Box 118, S-22100 Lund, Sweden}

\begin{abstract}

We study the ground state of few bosons with repulsive dipole-dipole interaction in a quasi-one-dimensional harmonic trap by means of the exact diagonalization method. Up to three interaction regimes are found depending on the strength of the dipolar interaction and the ratio of transverse to axial oscillator lengths: a regime where the dipolar Bose gas resembles a system of weakly $\delta$-interacting bosons, a second regime where the bosons are fermionized, and a third regime where the bosons form a Wigner crystal. In the first two regimes, the dipole-dipole potential can be replaced by a $\delta$ potential. In the crystalline state, the overlap between the localized wave packets is strongly reduced and all the properties of the boson system equal those of its fermionic counterpart. The transition from the Tonks-Girardeau gas to the solidlike state is accompanied by a rapid increase of the interaction energy and a considerable change of the momentum distribution, which we trace back to the different short-range correlations in the two interaction regimes.

\end{abstract}


\maketitle

{\it Note: This arXiv version contains at the end the Erratum to the published version~\cite{Deuretzbacher10}.}

\section{Introduction}

Ultracold atoms and molecules with large permanent dipole moments are currently attracting much interest, since they allow realization of quantum gas systems with long-range interactions. Major steps into this direction have already been done. References \cite{Lahaye07, Koch08} reported strong dipolar effects in a Bose-Einstein condensate (BEC) of $^{52}$Cr, which possesses a large permanent magnetic dipole moment of comparable strength as the usual $\delta$ interaction. Even more promising are ultracold molecules of two different atomic species, since they have much larger permanent electric dipole moments \cite{Aymar05}. They have already been produced by means of radio-frequency (rf) spectroscopy \cite{Ospelkaus06} and brought into the lowest internal vibrational ground state by means of a stimulated Raman adiabatic passage \cite{Ospelkaus08, Ni08}. What remains is to cool these gases down into the quantum degenerate regime.

Many new effects have been predicted for ultracold quantum gases with dipole-dipole interactions (DDIs), which are based on the long range and anisotropy of the DDI. Among others, we mention the stabilization of a dipolar BEC in a pancake-shape trap \cite{Santos00}, the roton-maxon character of the excitation spectrum \cite{Santos03}, new exotic quantum phases in optical lattices \cite{Goral02, Buechler07a, Buechler07b}, and the transfer of spin into angular momentum similar to the Einstein--de Haas effect in ferromagnets \cite{Santos06, Kawaguchi06}. Apart from that, the shape and strength of the intermolecular interactions may be controlled by means of static electric and microwave fields \cite{Buechler07a}, and even three-body interactions may be realized in optical lattices \cite{Buechler07b}. Moreover, novel quantum computation schemes are proposed with ultracold dipolar molecules \cite{DeMille02, Yelin06, Andre06}.

In this article, we discuss different interaction regimes of a quasi-one-dimensional (1D) dipolar Bose gas, in which the permanent dipole moments are aligned by an external field, such that the effective 1D DDI is repulsive. Thus far, theoretical studies of 1D dipolar bosons found correlations \cite{Arkhipov05, Citro07} beyond those in a Tonks-Girardeau (TG) gas \cite{Girardeau60, Kinoshita04, Kinoshita05} and excitations \cite{Pedri08, Astrakharchik08a, DePalo08}, which can be described within the Luttinger liquid framework \cite{Citro08}. We show that the $1 / |x|^3$ potential used in these references, which is valid in traps of zero transverse width, gives rise to an infinitely strong $\delta$ peak at equal particle positions, even for the infinitesimal strength of the DDI, due to its singular behavior at $x = 0$. Hence, the 1D Bose system forms a TG gas for very weak interactions \cite{Arkhipov05, Citro07, Pedri08}. We use a 1D DDI potential that accounts for the finite width of the trap and is hence finite at $x = 0$ \cite{Sinha07}. We show that this potential acts like a $\delta$ peak of finite strength, when the trap anisotropy is large. This allows for a regime of weakly $\delta$-interacting bosons below the TG regime. In the different interaction regimes, we analyze the contributions to the total energy, the particle density, the momentum, and occupation number distribution of the particles in the harmonic trap. Different features of the momentum distribution and the associated kinetic energy are related to the correlations between the particles.

\section{Model Hamiltonian}

\begin{figure}[t]
  \includegraphics[width = .5\columnwidth]{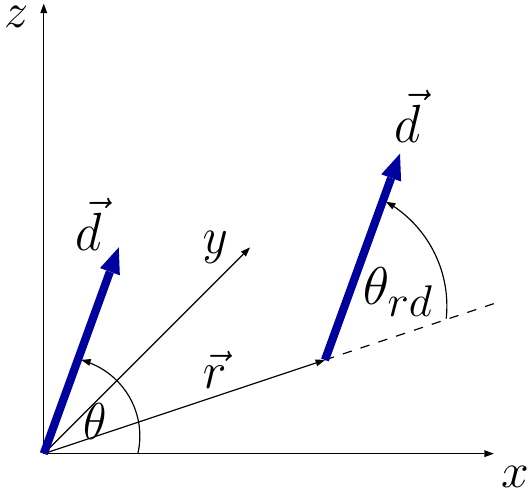}
  \caption{(Color online) Dipoles $\vec d$, which are oriented within the $x$-$z$ plane and which enclose an angle $\theta$ with the $x$ axis.}
  \label{Fig-sketch-of-dipoles}
\end{figure}

We consider electric or magnetic dipoles, which are aligned in the $xz$-plane by an external field (see Fig.~\ref{Fig-sketch-of-dipoles}). The DDI between two point-like dipoles is modeled by
\begin{equation} \label{Eq-3d-ddi}
  V_\text{dd} ( \vec r ) = \frac{d^2}{r^3} \bigl( 1 - 3 \cos^2 \theta_{rd} \bigr) ,
\end{equation}
where $d^2$ is the strength of the DDI and $\cos \theta_{rd} = \vec r \cdot \vec d / (r d)$. The coupling strength between two permanent electric dipole moments is given by $d^2 = (\widetilde{D} \, \text{debye})^2 / (4 \pi \epsilon_0)$, where $\widetilde{D}$ is the dipole moment in debyes and $\epsilon_0$ is the electric constant. In the case of permanent magnetic dipole moments, $d^2 = \mu_0 g_L^2 \mu_B^2 / (4 \pi)$, where $\mu_0$ is the magnetic constant, $g_L$ is the Land\'e factor, and $\mu_B$ is the Bohr magneton. In the following, we neglect the short-range van der Waals interaction \cite{Bortolotti06, Sinha07}.

\begin{figure}[b]
  \includegraphics[width = \columnwidth]{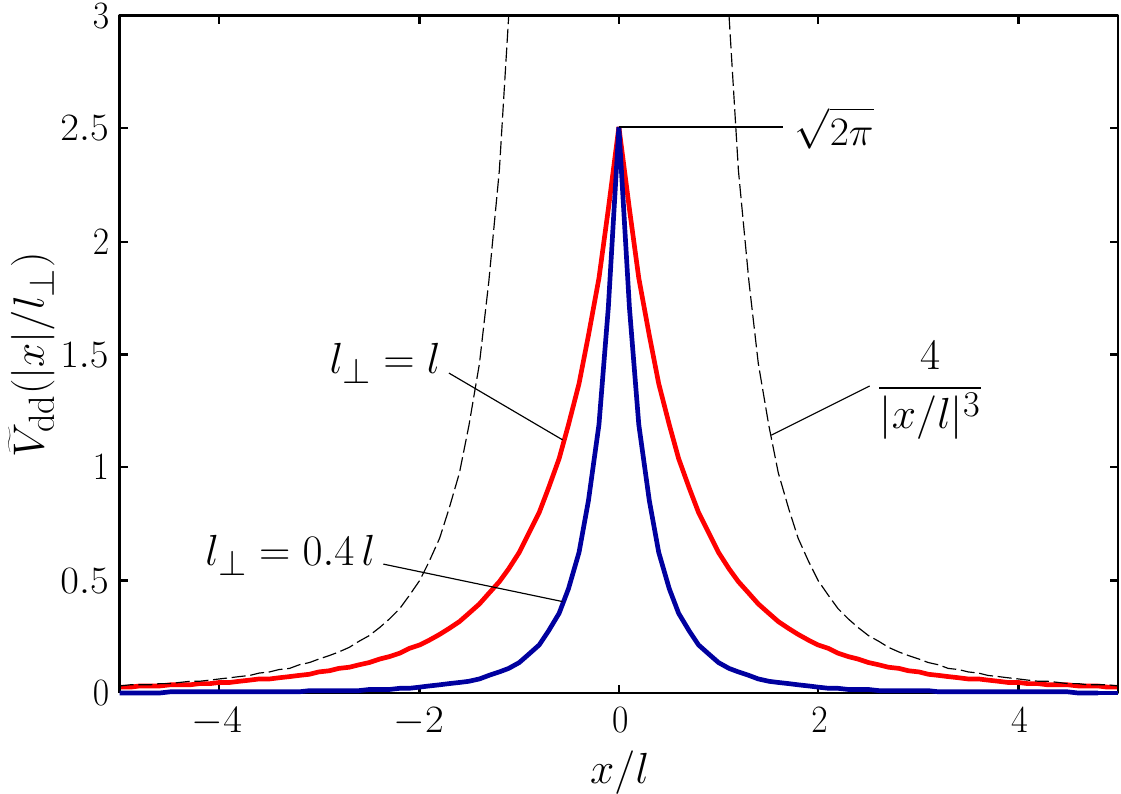}
  \caption{(Color online) Dimensionless DDI potential $\widetilde V_\text{dd} \bigl( |x| / l_\bot \bigr)$ for $l_\bot = 0.4 \, l$ (blue) and $l_\bot = l$ (red). $\widetilde V_\text{dd}$ is finite at the origin and becomes more peaked for smaller $l_\bot$. At large distances, $\widetilde V_\text{dd} \propto 1 / |x|^3$.}
  \label{Fig-dimensionless-1d-ddi}
\end{figure}

The dipoles are enclosed in a cigar-shaped harmonic trap, which may be generated by a deep optical lattice \cite{Kinoshita04}:
\begin{equation*}
  V_\text{trap} ( \vec r ) = \frac{m}{2} \Bigl[ \omega^2 x^2 + \omega_\bot^2 \bigl( y^2 + z^2 \bigr) \Bigr] .
\end{equation*}
Here, $\omega$ and $\omega_\bot$ are the trap frequencies of the axial and perpendicular directions, respectively, and $\omega \ll \omega_\bot$. Under the condition that the energy per particle of the axial direction is much smaller than the transverse level spacing $\hbar \omega_\bot$, one can assume that the particles stay in the ground state of the transverse harmonic oscillator, $\exp \bigl[ - (y^2 + z^2) / (2 l_\bot^2) \bigr] / (l_\bot \sqrt{\pi})$ with $l_\bot = \sqrt{\hbar / (m \omega_\bot)}$. Integration over the transverse directions yields the effective 1D DDI~\cite{Sinha07}
\begin{equation} \label{Eq-1d-ddi}
  V_\text{dd} ( x ) = U_\text{dd} \, \widetilde V_\text{dd} \bigl( |x| / l_\bot \bigr) ,
\end{equation}
with
\begin{equation} \label{Eq-Udd}
  U_\text{dd} = -\frac{d^2 [1 + 3 \cos (2 \theta)]}{8 \, l_\bot^3}
\end{equation}
and
\begin{equation} \label{Eq-dimensionless-1d-ddi}
  \widetilde V_\text{dd} (u) = - 2 u + \sqrt{2 \pi} \, \bigl( 1 + u^2 \bigr) \, e^{u^2 / 2} \, \text{erfc} \bigl( u / \sqrt{2} \bigr) ,
\end{equation}
where erfc is the complementary error function. An explicit calculation of Eqs.~(\ref{Eq-1d-ddi})--(\ref{Eq-dimensionless-1d-ddi}) is done in Appendix~\ref{Sec-calculation-of-the-1D-DDI}. The dimensionless DDI potential $\widetilde V_\text{dd} \bigl( |x| / l_\bot \bigr)$ is plotted in Fig.~\ref{Fig-dimensionless-1d-ddi}. The second-quantized many-particle Hamiltonian is then
\begin{equation} \label{Eq-Hamiltonian}
  H = \hbar \omega \sum_i \biggl( i + \frac{1}{2} \biggr) a_i^\dagger a_i + \frac{1}{2} \, U_\text{dd} \sum_{ijkl} \tilde{I}_{ijkl} a_i^\dagger a_j^\dagger a_l a_k \, ,
\end{equation}
where $a_i^\dagger$ $(a_i)$ are bosonic creation (annihilation) operators for one particle in energy eigenstate $\phi_i (x)$ of the axial harmonic oscillator and where
\begin{equation*}
  \tilde{I}_{ijkl} = \int_{-\infty}^\infty dx dx' \phi_i(x) \phi_j(x') \widetilde V_\text{dd} \bigl( |x - x'| / l_\bot \bigr) \phi_k(x) \phi_l(x')
\end{equation*}
are dimensionless interaction integrals. These integrals are integrated numerically for different $l_\bot$. The Hamiltonian matrix (\ref{Eq-Hamiltonian}) is diagonalized in the subspace of the energetically lowest eigenstates of the noninteracting many-particle problem.

\section{Discussion of the effective 1D DDI}
\label{discussion-of-the-1D-DDI}

We solve numerically the many-body Schr\"odinger equation with the interaction of Eqs.~(\ref{Eq-1d-ddi})--(\ref{Eq-dimensionless-1d-ddi}), but to gain some intuitive understanding let us first analyze the interaction potential. The DDI strength $U_\text{dd}$, as defined in (\ref{Eq-Udd}), is negative for angles between $0 \leqslant \theta < \theta_\text{crit.}$ with $\theta_\text{crit.} = \arccos (1 / \sqrt{3})$ and positive for $\theta_\text{crit.} < \theta \leqslant \pi / 2$. For $\theta = 0$, the 1D DDI is maximally attractive with $U_\text{dd} = -d^2 / (2 \, l_\bot^3)$, and for $\theta = \pi / 2$, it is maximally repulsive with $U_\text{dd} = d^2 / (4 \, l_\bot^3)$. In the following, we restrict the discussion to the repulsive case $U_\text{dd} > 0$.

From a Taylor expansion of (\ref{Eq-dimensionless-1d-ddi}) around infinity, one finds that $\widetilde V_\text{dd} (u) \rightarrow 4 / u^3$ for $u \rightarrow \infty$. Thus, for large distances $|x| \gg l_\bot$, the effective 1D DDI is given by
\begin{equation*}
  V_\text{dd} ( x ) \approx U_\text{lr} / |x / l|^3 ,
\end{equation*}
with the long-range interaction strength $U_\text{lr} = 4 \lambda^3 U_\text{dd}$, where $\lambda = l_\bot / l$. On the other hand, Fig.~\ref{Fig-dimensionless-1d-ddi} suggests that $V_\text{dd} ( x )$ becomes a $\delta$ peak for small $l_\bot$. Indeed one finds
\begin{equation*}
  \int_{-\infty}^\infty dx \frac{1}{4 \, l_\bot} \widetilde V_\text{dd} \bigl( |x| / l_\bot \bigr) = 1 ,
\end{equation*}
which leads us to the definition $\delta_{l_\bot} = \widetilde V_\text{dd} \bigl( |x| / l_\bot \bigr) / (4 \, l_\bot)$. We define the width of the $\delta_{l_\bot}$ function according to
\begin{equation*}
  w_{l_\bot} = 2 \sqrt{2 \pi} \, l_\bot \approx 5 \, l_\bot ,
\end{equation*}
which is justified by the observation that
\begin{equation*}
  \int_{-\sqrt{2 \pi} \, l_\bot}^{\sqrt{2 \pi} \, l_\bot} dx \frac{1}{4 \, l_\bot} \widetilde V_\text{dd} \bigl( |x| / l_\bot \bigr) \approx 90 \% .
\end{equation*}
Hence, the series of $\delta_{l_\bot}$ functions converges toward a $\delta$ peak when $l_\bot$ approaches zero; that is, ${\delta_{l_\bot} \rightarrow \delta}$ for ${l_\bot \rightarrow 0}$. We conclude that at short distances $|x| \lesssim 2.5 \, l_\bot$
\begin{equation*}
  V_\text{dd} ( x ) \approx 4 \, l_\bot U_\text{dd} \, \delta (x) = U_\text{sr} \, \widetilde \delta (x) ,
\end{equation*}
with $U_\text{sr} = 4 \lambda U_\text{dd}$ and $\delta = \widetilde \delta / l$. It follows that the strengths of the short- and long-range parts of the 1D DDI scale differently with the trap anisotropy $1 / \lambda$,
\begin{equation*}
  U_\text{sr} = U_\text{lr} / \lambda^2 ,
\end{equation*}
which means that the short-range part of the interaction is strongly enhanced by a tight transverse confinement compared to the long-range part. Particularly in the limit $\lambda = 0$, the effective 1D DDI is not only given by $U_\text{lr} / |x/l|^3$ at $x \neq 0$, but there is an additional infinitely strong $\delta$ peak at $x = 0$. Consider, for example, the situation of a rather small $U_\text{lr} = 0.1 \, \hbar \omega$ and a trap anisotropy of $1 / \lambda = 10$. Then, the strength of the $\delta$ peak is $U_\text{sr} = 10 \, \hbar \omega$, which is already so large that bosons form a TG gas.

This decomposition of the 1D DDI is valid as long as the ground state is a slowly varying function along the axial direction on the transverse length scale $l_\bot$. In the following we fix the trap anisotropy according to $1 / \lambda = 50$ so that this condition is always fulfilled. This corresponds to the maximal trap anisotropy, which has been achieved in the experiment of Kinoshita {\it et al.} \cite{Kinoshita04}, where $\omega = 2 \pi \, 27.5 \,$Hz and $\text{max}(\omega_\bot) = 2 \pi \, 70.7 \,$kHz leading to max$(1 / \lambda) = 50.7$.

In Sec. \ref{results}, we perform a sweep of $U_\text{dd}$, which can be done by changing the angle $\theta$ between $\arccos (1 / \sqrt{3})$ and $\pi / 2$. As mentioned before, we neglected the short-range van der Waals interaction, since we are interested in a study of the effect of the 1D DDI alone. These relations, which we derived within the single-mode approximation, are true as long as $l_\text{dd} \gtrsim l_\bot$ \cite{Sinha07}, where $l_\text{dd} = d^2 m / \hbar^2$. By taking $^{40}$K$^{87}$Rb polar molecules with an electric dipole moment of $0.6 \,$debye and $\omega_\bot = 2 \pi \, 70.7 \,$kHz, the condition is fulfilled for $\theta \gtrsim 1$. However, weak dipole moments may largely influence the 1D scattering in the regime $l_\bot \gg l_\text{dd}$, which was shown in a theory that accounts for the short-range van der Waals interaction and goes beyond the single-mode approximation \cite{Sinha07}. To study the effect of the 1D DDI alone in that regime, one may tune $g_\text{1D}$ to zero using a Feshbach resonance for fixed $\theta$ and $\omega_\bot$ and change $\omega$. An increase of $\omega$, for example, would decrease $U_\text{sr} / (\hbar \omega)$ and increase $U_\text{lr} / (\hbar \omega)$.

\section{Results}
\label{results}

\begin{figure}[t]
  \includegraphics[width = \columnwidth]{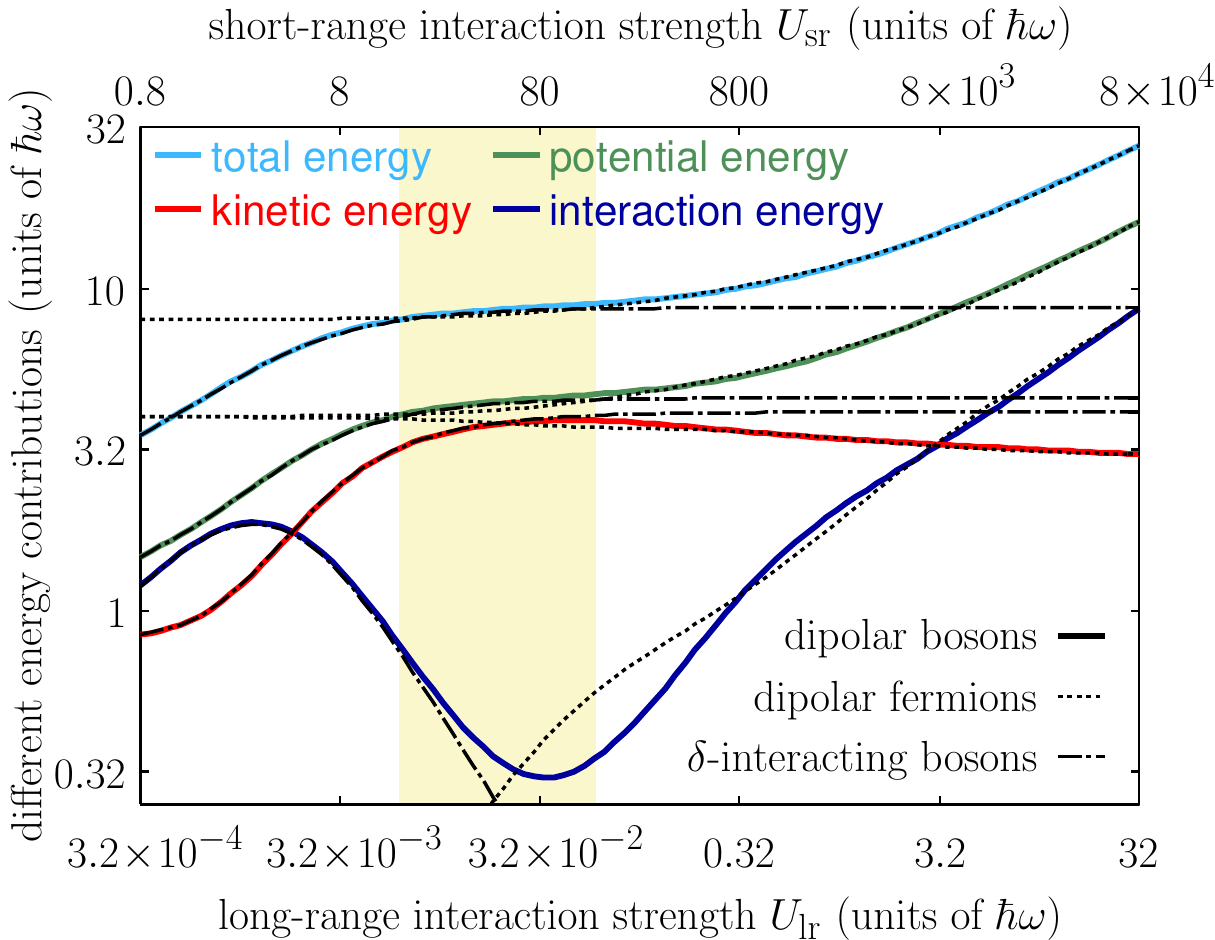}
  \caption{(Color online) Contributions to the total energy of four particles as a function of $U_\text{lr}$ (lower scale) and $U_\text{sr}$ (upper scale), respectively, in a double logarithmic plot for a trap anisotropy of $1 / \lambda = 50$. Dipolar bosons (solid) are compared to $\delta$-interacting bosons (dash-dotted) and dipolar fermions (dotted). The shaded background marks the TG regime, where the bosons fermionize.}
  \label{Fig-energy-contributions}
\end{figure}

Let us now turn to our results for four particles obtained by means of a numerical diagonalization of (\ref{Eq-Hamiltonian}). Figure~\ref{Fig-energy-contributions} shows the contributions to the total energy as a function of $U_\text{lr}$ (lower scale) and $U_\text{sr}$ (upper scale), respectively, in a double logarithmic plot for a trap anisotropy of $1 / \lambda = 50$. Both scales are related to each other through $U_\text{sr} = U_\text{lr} / \lambda^2$. Shown are the energies of bosons with DDI (solid) and $\delta$ interaction (dash-dotted) and of fermions with DDI (dotted). One sees that all energy contributions of bosons with DDI agree with those of $\delta$-interacting bosons for $U_\text{lr} \lesssim 0.1 \, \hbar \omega$; that is, in this region the bosons feel only the short-range $\delta$ part of the DDI. Hence, it is more illuminative to use the upper scale here. The transition behavior in this interaction regime has been discussed in Ref.~\cite{Deuretzbacher07} (for other trap geometries, see \cite{Hao06, Zoellner06}): The system evolves from a weakly interacting quasi-BEC via an intermediate regime to a TG gas. Already at $U_\text{sr} \approx 10 \, \hbar \omega$, the bosons are fermionized, which is indicated by a saturation of the total energy. In the shaded region, the system does not react to a further increase of the interaction strength, which shows that the bosons do not feel the long-range part of the DDI, which is of the order of $U_\text{lr}$ at distance $l$. However, when the DDI is increased above the critical value $U_\text{lr} \gtrsim 0.1 \, \hbar \omega$, the fermionized bosons are further pushed apart from each other by the long-range $1 / |x|^3$ tail of the DDI. An obvious signature of these beyond-TG correlations is the rapid increase of the interaction energy, indicating that the $1 / |x|^3$-tail of the DDI has significant overlap with the many-body wave function. Clearly, since the bosons are fermionized, all the energy contributions coincide with those of fermions with DDI in this region.

\begin{figure}[t]
  \includegraphics[width = \columnwidth]{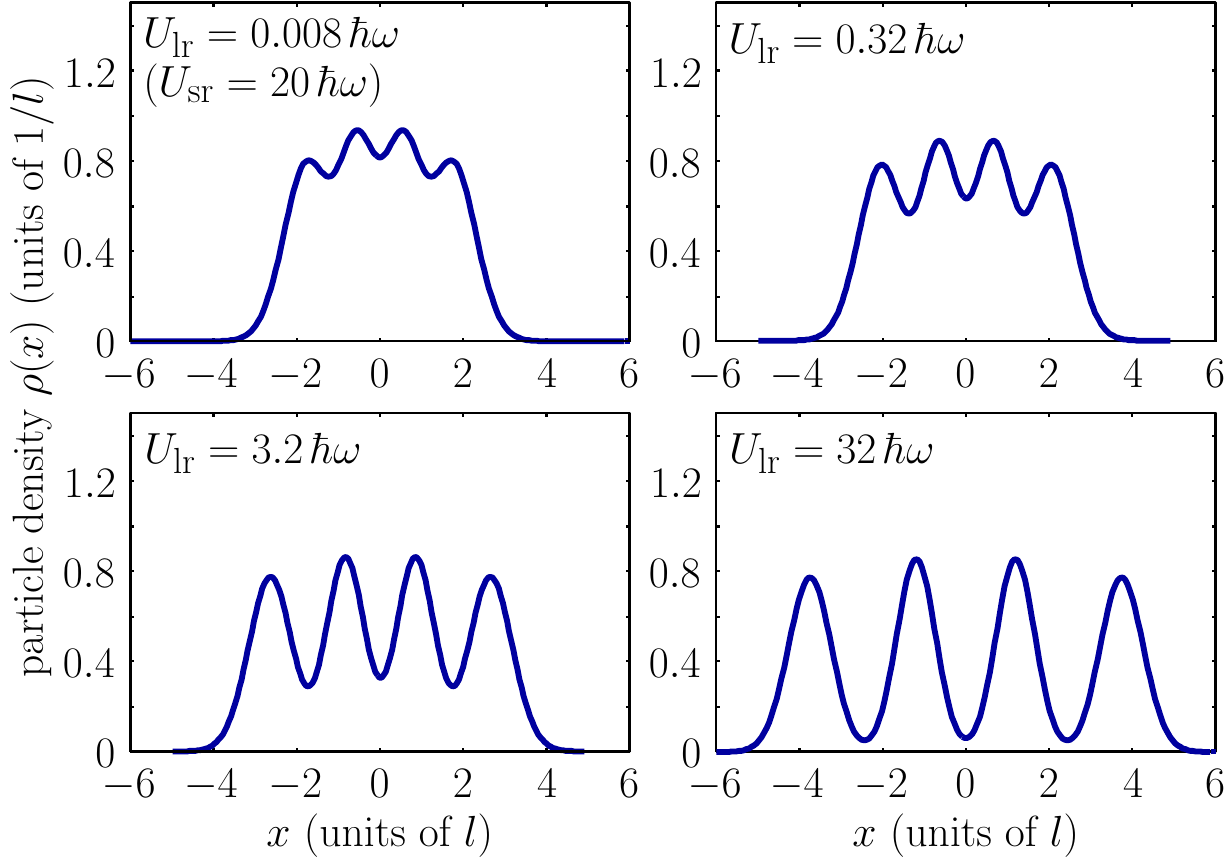}
  \caption{(Color online) Density of four particles for different interaction strengths. With increasing repulsion, the particles localize due to the long range of the interaction. For the shown values of $U_\text{lr}$, the densities of fermions and bosons are equal, since the bosons are fermionized.}
  \label{Fig-density}
\end{figure}

Another interesting aspect concerns the kinetic energy, which decreases in the $1 / |x|^3$-tail-dominated regime, while all the other energy contributions grow with increasing repulsion. We attribute this behavior to the short-range correlations of the wave function, which can be approximated by \cite{DePalo08}
\begin{equation} \label{Eq-correlated-trial-wf}
  \psi (x_1, \ldots, x_N) \propto \prod_{i < j} |x_i - x_j|^{1 / K} \prod_k e^{- x_k^2 / (2 l)} .
\end{equation}
In the TG regime, the Luttinger exponent is $K = 1$, which leads to a large gradient of the many-body wave function at $x_i = x_j$. This gives rise to a rather large kinetic energy in these regions of the configuration space. With increasing long-range interactions, the Luttinger exponent decreases, $K < 1$ \cite{Citro07}, which diminishes the gradient and hence the kinetic energy at $x_i = x_j$. For even stronger long-range interactions, the many-body wave function can alternatively be approximated by localized Gaussian-like wave packets \cite{Arkhipov05, DePalo08}. As can be seen in Fig.~\ref{Fig-density}, the repulsion between the bosons mainly increases the distance between the center points of the wave packets, which has no influence on the kinetic energy. Hence, we expect a saturation of the kinetic energy for very strong repulsion, which eventually becomes negligible compared to the other energy contributions; see Fig.~\ref{Fig-energy-contributions}.

The main message of Fig.~\ref{Fig-energy-contributions} is the distinction of three interaction regimes. In the left region, the bosons feel a $\delta$ potential of finite strength; in the middle region, the bosons feel an infinitely strong $\delta$ potential and thus fermionize; and in the right region, the system is dominated by the long-range $1 / |x|^3$ part of the DDI. The right boundary of the middle region is independent of the trap anisotropy $1 / \lambda$, but the left boundary moves to the left when $1 / \lambda$ is increased, and in the limit $1 / \lambda = \infty$, the left region is absent. On the other hand, for very low trap anisotropies, the width of the middle region shrinks to zero. In Fermi systems, the left interaction regime is absent, since fermions do not feel the $\delta$ part of the DDI.

\begin{figure}[t]
  \includegraphics[width = \columnwidth]{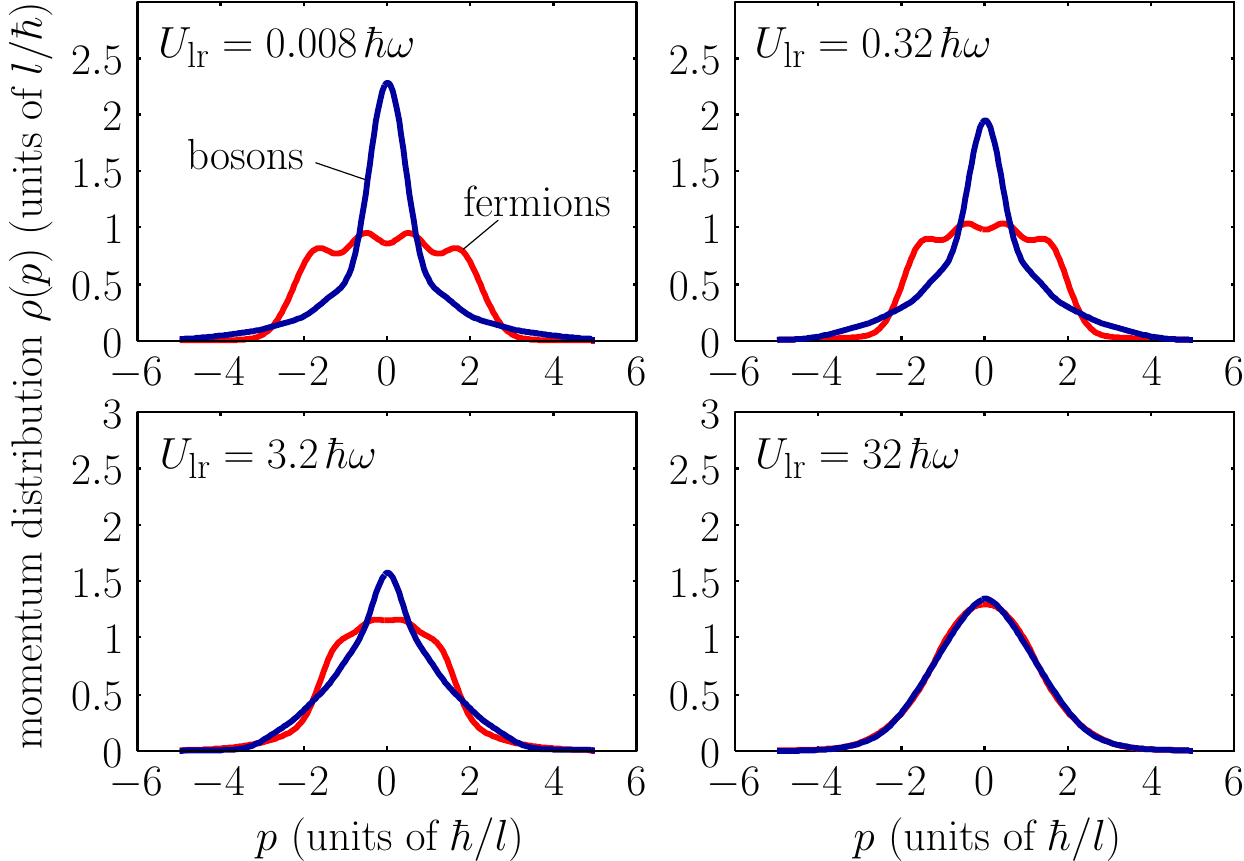}
  \caption{(Color online) Momentum distribution of four particles for different strengths of $U_\text{lr}$. The top left figure shows the momentum distribution of noninteracting fermions and fermionized bosons, which are different. For very strong repulsion, the distributions of fermions and bosons become equal. For $U_\text{lr} = 32 \, \hbar \omega$, the momentum distributions are the Fourier transform of the localized wave packets of the bottom right particle density of Fig.~\ref{Fig-density}.}
  \label{Fig-momentum-distribution}
\end{figure}

Since the two $\delta$-interaction-dominated regimes have already been discussed in Ref.~\cite{Deuretzbacher07}, we concentrate in the following on the $1 / |x|^3$-tail-dominated regime. Figure~\ref{Fig-density} shows localization of the particles with increasing long-range interaction. In the TG regime, the particles are rather close together, and the minor oscillations visible in the top left density of Fig.~\ref{Fig-density} disappear for large particle numbers $N$. With increasing long-range interaction, the density resembles four overlapping localized wave packets, which move apart from each other. The equilibrium positions of the wave packets minimize the potential and interaction energy (the kinetic energy is negligible). Occurrence of this quasiordered state was identified in the homogeneous system by means of the correlation function \cite{Arkhipov05} and the static structure factor \cite{Citro07}.

Particularly interesting is the momentum distribution (Fig.~\ref{Fig-momentum-distribution}), which is different for bosons and fermions in the TG regime (in contrast to the energies and densities). One sees that both distributions become equal for strong long-range repulsion. This reveals that the statistics of the particles become unimportant if there is no significant overlap between the wave packets of the individual particles.

In the TG regime (Fig.~\ref{Fig-momentum-distribution}, top left), the bosonic momentum distribution exhibits a high zero-momentum peak and long-range high-momentum tails (this is more clearly visible in Ref. \cite{Deuretzbacher07}). The high-momentum tails originate from the cusps in the TG wave function at $x_i = x_j$ \cite{Minguzzi02, Olshanii03}. Note that although the bosonic and fermionic distributions are markedly different in the TG regime, their kinetic energy $E_\text{kin.} = \int dp p^2 \rho(p) / (2 m)$ is equal. This shows that the high-momentum tails (and hence the short-range correlations) contribute significantly to the kinetic energy.

With increasing repulsion, the zero-momentum peak of the bosonic distribution significantly decreases and the high-momentum tails vanish, which indicates a redistribution from low and high toward medium momenta. Finally, for very strong repulsion (Fig.~\ref{Fig-momentum-distribution}, bottom right), the bosonic and fermionic distributions are only marginally different. The broad Gaussian distribution of the crystalline state is essentially the Fourier transform of the wave packets of Fig.~\ref{Fig-density}, like in the Mott insulator phase in an optical lattice \cite{Greiner02}.

\begin{figure}[t]
  \includegraphics[width = \columnwidth]{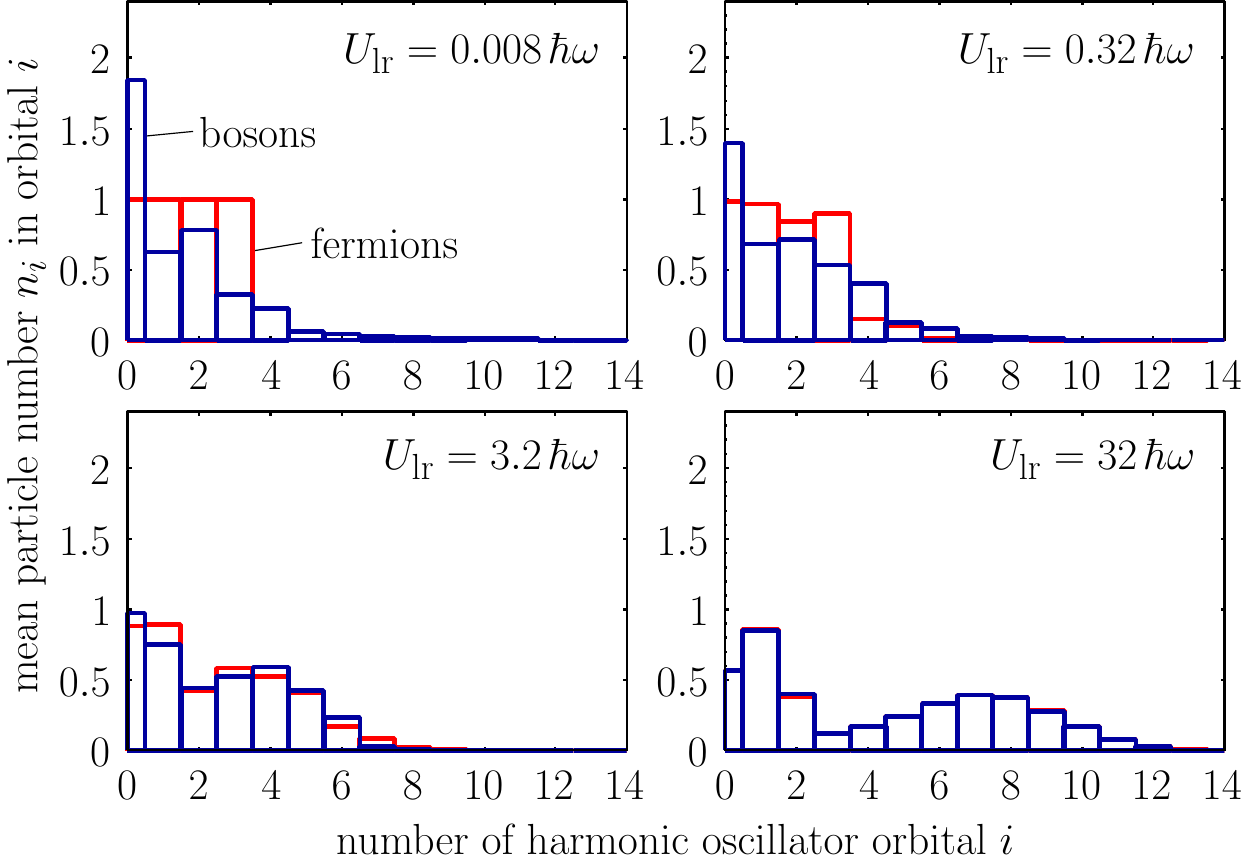}
  \caption{(Color online) Occupation number distribution of four particles in the 1D harmonic oscillator states for different values of $U_\text{lr}$. The top left figure shows clear differences between fermionized bosons and noninteracting fermions. With increasing repulsion, the distributions of fermions and bosons become equal.}
  \label{Fig-occupation-number-distribution}
\end{figure}

We close our discussion with Fig.~\ref{Fig-occupation-number-distribution}, which shows the occupation number distribution of the 1D harmonic oscillator states. As in the case of the momentum distribution, one sees that the bosonic and fermionic distributions become equal for strong long-range repulsion. For $U_\text{lr} = 32 \, \hbar \omega$, the distribution exhibits two features: a sharp peak with a maximum population of the first excited harmonic oscillator state and a broad distribution over the states $i = 3 - 14$ with a maximum at $i = 7$. The broad maximum shifts to larger $i$ when $U_\text{lr}$ is further increased.

\section{Conclusions}

We analyzed several microscopic ground-state properties of few dipolar bosons in a 1D harmonic trap. The interaction between the bosons was modeled by a 1D potential, which was obtained from the three-dimensional DDI potential by an integration over the transverse directions within the single-mode approximation. For large trap anisotropies, the 1D DDI acts on the ground state like the sum of a short-range $\delta$ and a long-range $1 / |x|^3$ potential. Depending on the relative strength of both contributions, the system forms a weakly interacting quasi-BEC, a TG gas, or a solidlike state of particles, which are localized due to the strong long-range repulsion.

The transition from the TG gas to the solidlike state was clearly visible in the momentum distribution. With increasing interaction, one observes a redistribution from low and high toward medium momenta. The disappearance of the high-momentum tails (and hence the short-range correlations) is responsible for the decrease of the kinetic energy. For very strong repulsion, the momentum distribution has a broad Gaussian shape and is essentially the Fourier transform of the localized wave packets. In contrast to the kinetic energy, the interaction and the potential energy grow rapidly with increasing repulsion in the solidlike state.

We remark, that for weak dipolar interactions one has to consider scattering effects beyond the single-mode approximation \cite{Sinha07}; see the discussion at the end of Sec.~\ref{discussion-of-the-1D-DDI}. Hence, one may use Feshbach resonances in order to enter the regime of weak interactions in a controlled way.

\begin{acknowledgments}

We thank L. Santos, S. Ospelkaus, D. Pfannkuche, and G. Kavoulakis for valuable discussions. This work was financially supported by the Swedish Research Council and the Swedish Foundation for Strategic Research. F.~D. acknowledges funding by the DFG (SFB407, QUEST), the ESF (EUROQUASAR), and the hospitality of S.~M.~R. during his stay at the LTH, where a large part of this research was performed.

\end{acknowledgments}

\begin{appendix}

\section{Calculation of the 1D DDI}
\label{Sec-calculation-of-the-1D-DDI}

It is assumed that the particles reside in the ground state of the transverse directions. The effective 1D DDI is given by
\begin{eqnarray*}
  V_\text{dd} ( x_1 - x_2 ) & = & \int dy_1 dy_2 dz_1 dz_2 V_\text{dd} ( \vec r_1 - \vec r_2 ) \\
  & & \times \, \phi_0^2 (y_1) \phi_0^2 (y_2) \phi_0^2 (z_1) \phi_0^2 (z_2) ,
\end{eqnarray*}
with $\phi_0 (u) = \exp \bigl[ -u^2 / (2 l_\bot^2) \bigr] / \sqrt{l_\bot \sqrt{\pi}}$ and $V_\text{dd} ( \vec r_1 - \vec r_2 )$ given by Eq.~(\ref{Eq-3d-ddi}). We introduce relative and center-of-mass coordinates $\vec r_{1/2} = \vec R \pm \vec r / 2$ and perform the integration over $Y$ and $Z$:
\begin{equation*}
  V_\text{dd} (x) = \frac{1}{2 \pi l_\bot^2} \int dy dz V_\text{dd} ( \vec r ) e^{-(y^2 + z^2) / (2 l_\bot^2)} .
\end{equation*}
In cylindrical coordinates $\vec r = (x, y, z) = (x, \rho \cos \phi, \rho \sin \phi)$ the integral becomes
\begin{equation*}
  V_\text{dd} (x) = \frac{1}{2 \pi l_\bot^2} \int d\phi d\rho V_\text{dd} ( x, \rho, \phi ) \rho \, e^{-\rho^2 / (2 l_\bot^2)} ,
\end{equation*}
where $V_\text{dd} ( x, \rho, \phi )$ is given by
\begin{equation*}
  V_\text{dd} ( x, \rho, \phi ) = \frac{d^2}{\sqrt{x^2 + \rho^2}^3} \bigl( 1 - 3 \cos^2 \theta_{rd} \bigr) ,
\end{equation*}
with
\begin{equation*}
  \cos \theta_{rd} = \frac{\vec r \cdot \vec d}{r d} = \frac{x \cos \theta + \rho \sin \phi \sin \theta}{\sqrt{x^2 + \rho^2}} .
\end{equation*}
Here, we assumed that the dipoles lie in the $x-z$ plane (see Fig.~\ref{Fig-sketch-of-dipoles}), so that $\vec d / d = ( \cos \theta, 0, \sin \theta)$. Integration of $\cos^2 \theta_{rd}$ over $\phi$ gives
\begin{equation*}
  \int_0^{2 \pi} d\phi \cos^2 \theta_{rd} = \frac{\pi}{x^2 + \rho^2} \bigl( 2 x^2 \cos^2 \theta + \rho^2 \sin^2 \theta \bigr) .
\end{equation*}
Hence, we obtain
\begin{equation*}
  \frac{1}{2 \pi l_\bot^2} \int_0^{2 \pi} d\phi V_\text{dd} ( x, \rho, \phi ) = A_1 \frac{\rho^2 - 2 x^2}{\sqrt{x^2 + \rho^2}^5}
\end{equation*}
with $A_1 = d^2 [ 1 + 3 \cos (2 \theta) ] / (4 l_\bot^2)$. It remains to perform the integration over $\rho$
\begin{equation*}
  V_\text{dd} (x) = A_1 \int_0^\infty d\rho \rho \, e^{-\rho^2 / (2 l_\bot^2)} \frac{\rho^2 - 2 x^2}{\sqrt{x^2 + \rho^2}^5} .
\end{equation*}
We substitute $u^2 = x^2 + \rho^2$ $(\rightarrow u du = \rho d\rho)$ and obtain
\begin{equation*}
  V_\text{dd} (x) = A_1 e^{x^2 / (2 l_\bot^2)} \int_{|x|}^\infty du \, e^{-u^2 / (2 l_\bot^2)} \frac{u^2 - 3 x^2}{u^4} .
\end{equation*}
Next, we set $v = u / (\sqrt{2} l_\bot)$, which leads to
\begin{equation*}
  V_\text{dd} (x) = A_2 e^{\alpha^2} \int_\alpha^\infty dv \, e^{-v^2} \frac{2 l_\bot^2 v^2 - 3 x^2}{v^4} ,
\end{equation*}
with $\alpha = |x| / (\sqrt{2} l_\bot)$ and $A_2 = d^2 [ 1 + 3 \cos (2 \theta) ] / (8 \sqrt{2} l_\bot^5)$. From an integration by parts, we obtain the recurrence relation
\begin{equation*}
  I_n = \int_\alpha^\infty dv \frac{e^{-v^2}}{v^n} = \frac{e^{-\alpha^2}}{(n - 1) \alpha^{n - 1}} - \frac{2}{n - 1} I_{n - 2} ,
\end{equation*}
which is valid for $n = 2, 4, 6, \ldots \;$. Since $I_0 = \sqrt{\pi} \, \text{erfc} (\alpha) / 2$, the integrals $I_2$ and $I_4$ are given by
\begin{equation*}
  I_2 = \frac{e^{-\alpha^2}}{\alpha} - \sqrt{\pi} \, \text{erfc} (\alpha)
\end{equation*}
and
\begin{equation*}
  I_4 = \frac{e^{-\alpha^2}}{3 \alpha^3} + \frac{2}{3} \biggl[ \sqrt{\pi} \, \text{erfc} (\alpha) - \frac{e^{-\alpha^2}}{\alpha} \biggr] .
\end{equation*}
This is inserted into
\begin{equation*}
  V_\text{dd} (x) = A_2 e^{\alpha^2} \bigl( 2 l_\bot^2 I_2 - 3 x^2 I_4 \bigr) ,
\end{equation*}
which becomes
\begin{eqnarray*}
  V_\text{dd} (x) & = & A_2 \biggl( \frac{2 l_\bot^2}{\alpha} - 2 \sqrt{\pi} l_\bot^2 e^{\alpha^2} \text{erfc} (\alpha) - \frac{x^2}{\alpha^3} \\
  & & \mspace{32mu} - 2 \sqrt{\pi} x^2 e^{\alpha^2} \text{erfc} (\alpha) + \frac{2 x^2}{\alpha} \biggr) .
\end{eqnarray*}
Using the definition of $\alpha$, one sees, that the first and the third term cancel each other, and one obtains
\begin{equation*}
  V_\text{dd} (x) = A_2 \sqrt{2} \Bigl[ 2 l_\bot |x| - \sqrt{2 \pi} \bigl( l_\bot^2 + x^2 \bigr) e^{\alpha^2} \text{erfc} (\alpha) \Bigr] ,
\end{equation*}
which equals (\ref{Eq-1d-ddi})--(\ref{Eq-dimensionless-1d-ddi}).

\end{appendix}

\bibliographystyle{prsty}

\clearpage

\widetext

\section*{\large Erratum: Ground-state properties of few dipolar bosons in a quasi-one-dimensional harmonic trap}

In the Appendix of our paper, we derive the effective one-dimensional (1D) dipole-dipole interaction (DDI) potential. The derivation and the result are incomplete, since the singularity of the three-dimensional (3D) DDI at zero distance is not properly accounted for. To see this, we introduce a small $\epsilon > 0$ in the denominator of the 3D DDI. The first steps of the calculation are analogous to those shown in the Appendix until one arrives at~\footnote{Equation numbers in this erratum do not reflect the numbering in the original paper.}
\begin{equation}
V_\text{dd}^{(\epsilon)} (u) = \frac{1}{4} \biggl[ 1 + 3 \cos (2 \theta) \biggr] \frac{d^2}{l_\bot^3} \int_0^\infty dw \, w \, e^{-w^2/2} \frac{w^2 - 2 u^2}{\sqrt{w^2 + u^2 + \epsilon^2}^5}
\end{equation}
with $w = \rho / l_\bot$ and $u = x / l_\bot$. This is equal to
\begin{eqnarray}
V_\text{dd}^{(\epsilon)} (u) & = & \frac{1}{4} \biggl[ 1 + 3 \cos (2 \theta) \biggr] \frac{d^2}{l_\bot^3} \left\{ \int_0^\infty dw \, w \, e^{-w^2/2} \frac{w^2 - 2 (u^2 + \epsilon^2)}{\sqrt{w^2 + (u^2 + \epsilon^2)}^5} + 2 \epsilon^2 \int_0^\infty dw \, w \frac{e^{-w^2/2}}{\sqrt{w^2 + u^2 + \epsilon^2}^5} \right\} \nonumber \\
& = & -\frac{1}{8} \biggl[ 1 + 3 \cos (2 \theta) \biggr] \frac{d^2}{l_\bot^3} \Biggl\{ \biggl[ - 2 v + \sqrt{2 \pi} \, \bigl( 1 + v^2 \bigr) \, e^{v^2 / 2} \, \text{erfc} \bigl( v / \sqrt{2} \bigr) \biggr] - \frac{8}{3} \delta_\epsilon (u) \Biggr\}
\end{eqnarray}
with $v = \sqrt{u^2 + \epsilon^2}$ and
\begin{equation}
\delta_\epsilon (u) = \frac{3}{2} \int_0^\infty dw \, w \frac{\epsilon^2 e^{-w^2/2}}{\sqrt{w^2 + u^2 + \epsilon^2}^5} .
\end{equation}
In the limit $\epsilon \rightarrow 0$, one finally obtains
\begin{equation} \label{vdd}
\lim_{\epsilon \rightarrow 0} V_\text{dd}^{(\epsilon)} (u) = -\frac{1}{8} \biggl[ 1 + 3 \cos (2 \theta) \biggr] \frac{d^2}{l_\bot^3} \Biggl\{ \biggl[ - 2 u + \sqrt{2 \pi} \, \bigl( 1 + u^2 \bigr) \, e^{u^2 / 2} \, \text{erfc} \bigl( u / \sqrt{2} \bigr) \biggr] - \frac{8}{3} \delta (u) \Biggr\} ,
\end{equation}
i.~e., apart from the 2nd term in square brackets, which is already given in the paper, there is an additional $\delta$ term present in the 1D DDI. This term is missing in Eqs. (\ref{Eq-1d-ddi})--(\ref{Eq-dimensionless-1d-ddi}) of our paper. Let us briefly show that $\lim_{\epsilon \rightarrow 0} \int_{-\infty}^\infty du \, \delta_\epsilon (u) = 1$. One finds that
\begin{eqnarray}
\int_{-\infty}^\infty du \, \delta_\epsilon (u) & = & \frac{3}{2} \int_0^\infty dw \, \epsilon^2 w e^{-w^2/2} \left[ \frac{u ( 2 u^2 + 3 (w^2 + \epsilon^2) )}{3 (w^2 + \epsilon^2)^2 \sqrt{w^2 + u^2 + \epsilon^2}^3} \right]_{u = -\infty}^\infty = 2 \epsilon^2 \int_0^\infty dw \frac{w e^{-w^2/2}}{(w^2 + \epsilon^2)^2} \nonumber \\
& = & \frac{1}{2} \epsilon^2 e^{\epsilon^2/2} \int_{\epsilon^2/2}^\infty dt \frac{e^{-t}}{t^2} = 1 + \frac{1}{2} \epsilon^2 e^{\epsilon^2/2} \, \text{Ei} \left( -\epsilon^2/2 \right) \rightarrow 1 \quad \text{for} \; \epsilon \rightarrow 0 .
\end{eqnarray}
The 3rd step follows from the substitution $t = (w^2 + \epsilon^2)/2$, the 4th step from an integration by parts, and the last step from $\text{Ei}(x) \approx \ln(x)$ and $x^2 \ln(x) \rightarrow 0$ for $x \rightarrow 0$, where $\text{Ei}(x)$ is the exponential integral.

It can be seen from Eq.~(\ref{vdd}) that the strength of the $\delta$ interaction depends on the dipole orientation with respect to the weak trap axis (the angle $\theta$). This is different from the $\delta$ contribution, which stems from the point limit of a real (extended) dipole~\cite{Griffiths82}, where the strength depends on the relative orientation of the two interacting dipoles. In total there are three $\delta$ terms, which originate from the van der Waals interaction, the point limit of a real dipole and the integration over the transverse directions.

In our paper, we study the effect of the 1D DDI without the $\delta$ terms and perform a sweep of the interaction strength. Such a sweep can be performed experimentally, when the strength of all $\delta$ terms is tuned to zero by means of a Feshbach resonance. Then, the strength of the interaction term in square brackets with respect to the level spacing can be tuned by changing the axial angular frequency, as described at the end of Sec.~\ref{discussion-of-the-1D-DDI}.

As a test of Eq.~(\ref{vdd}) we finally calculate the interaction energy of $N$ bosonic dipoles, which are oriented along the $x$-axis $(\theta = 0)$ and occupy the ground state of the axial harmonic oscillator $\phi_0 (x) = e^{-x^2/(2 l^2)} / \sqrt{l \sqrt{\pi}}$. The result is given by
\begin{equation}
E_\text{int} = -\frac{N^2 d^2}{3 \sqrt{2 \pi} l l_\perp^2} \left[ \frac{1 + 2 \kappa^2}{1 - \kappa^2} - \frac{3 \kappa^2 \mathrm{artanh} \left( \sqrt{1 - \kappa^2} \right)}{\sqrt{1 - \kappa^2}^3} \right]
\end{equation}
with $\kappa = l_\perp / l$. This agrees with the 2nd term of Eq.~(5.8) in the review of T. Lahaye {\it et al.}~\cite{Lahaye09} [note that $a_\text{dd} = m C_\text{dd} / (12 \pi \hbar^2)$, $C_\text{dd} = 4 \pi d^2$, $a_\text{ho} = \sqrt{\hbar / (m \bar \omega)}$, $\sigma_\rho = l_\rho / a_\text{ho}$ and $\sigma_z = l_z / a_\text{ho}$]. The function in square brackets is $1$ in the limit $\kappa \rightarrow 0$ and decreases monotonously to $-2$ in the limit $\kappa \rightarrow \infty$ with a zero-crossing at $\kappa = 1$. In a cigar-shaped trap $(\kappa \ll 1)$ the dipoles are mainly in a head-to-tail configuration, in which the DDI is attractive, and hence the interaction energy is negative, $E_\text{int} < 0$. In a pancake-shaped trap $(\kappa \gg 1)$ the dipoles are mainly in a side-by-side configuration, in which the DDI is repulsive, and hence the interaction energy is positive, $E_\text{int} > 0$. In marked contrast to this behavior, the interaction energy would always have been negative if the $\delta$ term in Eq.~(\ref{vdd}) had been neglected.

\bibliographystyle{prsty}

\end{document}